\documentclass[preprint,preprintnumbers, prd, floatfix, superscriptaddress,nofootinbib] {revtex4-1}
\usepackage{epsfig}
\usepackage{subfigure}
\usepackage{dcolumn}
\usepackage{bm}
\usepackage[usenames ,dvipsnames]{xcolor}
\usepackage{slashed}
\usepackage{graphicx,color}
\usepackage[bookmarksnumbered, pdfstartview=FitH,colorlinks,urlcolor=blue, citecolor=blue,linkcolor=blue] {hyperref}
\begin{document}
\title{Muon mass correction in partial wave analyses of charmed meson semi-leptonic decays}

\author{Han Zhang}
\affiliation{School of Physics and Microelectronics,
Zhengzhou University, Zhengzhou, Henan 450001, China}

\author{Bai-Cian Ke}
\email{Corresponding author: baiciank@ihep.ac.cn}
\affiliation{School of Physics and Microelectronics,
Zhengzhou University, Zhengzhou, Henan 450001, China}

\author{Yao Yu}
\email{Corresponding author: yuyao@cqupt.edu.cn}
\affiliation{Chongqing University of Posts \& Telecommunications, Chongqing, 400065, China}

\author{En Wang}
\affiliation{School of Physics and Microelectronics,
Zhengzhou University, Zhengzhou, Henan 450001, China}

\date{\today}

\begin{abstract}
  We derive the parameterization formula for partial wave analyses of charmed
  meson semi-leptonic decays with consideration of the effects caused by the
  lepton mass. As the proposed super-tau-charm factory will reach much enhanced
  luminosity and BESIII is taking $\psi(3770)\to D\bar{D}$ data, our results
  are helpful to improve the measurement precision of future partial wave
  analyses of charmed meson semi-muonic decays. 
\end{abstract}

\maketitle
\section{introduction}
The Standard Model, as the theory of elementary particle interactions, has
achieved great success to describe experimental data with various energies.
Nevertheless, charmed mesons' $\sim 2$~GeV masses make them in the region
where perturbative QCD is not applicable, and raise challenges to both theory
and experiment. In recent years, testing the Standard Model with high
precision measurements become one of the hottest topics in the charm sector. 

The semi-leptonic decays of charmed mesons, in which the hadronic and weak
currents could be well separated, provide a clean platform to study the
mechanism of the $c$ quark to the $d(s)$ quark transition and play an
important role to understand the strong and weak interactions. Their partial
decay width accesses to the product of the hadronic form factor, which
describes the strong-interaction in the hadronic current connecting initial
and final hadrons, and the Cabibbo-Kobayashi-Maskawa matrix element $|V_{cs}|$
or $|V_{cd}|$, which parameterizes the weak interaction between different quark
flavors. Partial wave analyses of four-body semi-leptonic decays of charmed
meson allow to extract the form factors in the $D\to V \ell^+\nu_\ell$
and $D\to S \ell^+\nu_\ell$ transitions (where $\ell=e,\mu$, and $V$ and
$S$ denote vector and scalar mesons, respectively). The $K^{*}(892)^{-(0)}$
resonance have been studied in the $D^{0(+)}\to K^{0(-)}\pi^+e^+\nu_e$
decay by the CLEO, BABAR, and BESIII
collaborations~\cite{CLEO:2010enr, BaBar:2010vmf, BESIII:2015hty, BESIII:2018jjm}.
BESIII has also studied  $\rho^-$ resonance in
$D^{0}\to \pi^-\pi^0e^+\nu_e$~\cite{BESIII:2018qmf}; $\rho^0$, $\omega$,
and $f_0(500)$ resonances in
$D^{+}\to \pi^+\pi^-e^+\nu_e$~\cite{BESIII:2018qmf}. In the near future,
more partial wave analyses are expected to be performed with high-statistics
datasets, including $D_{(s)}$ semi-muonic decays.\footnote{At present, there is
no partial wave analyses of $D_{(s)}$ semi-muonic decays reported yet because
low statistics and high-level background caused by $\mu$-$\pi$
misidentification.}

However, the mass of leptons is neglected in the parameterization formula of
the amplitude analyses for the charmed meson semi-leptonic decays. This should
be fine in the case of semi-electronic decays, but will cause bias in
semi-muonic decays, which downgrades the advantage of high-statistics from the
new data and is against the purpose of precision measurement. In this work, we
derive the formula to consider the mass of leptons based on
Refs.~\cite{Pais:1968zza,Lee:1992ih}. The results will be present in the format
used in experimental analyses. Experiments can easily adopt our results to
their analyses. Charge-conjugated decay modes are implied throughout this
paper.

\section{Formalism}
\begin{figure}[t!]
\centering
\includegraphics[width=4.0in]{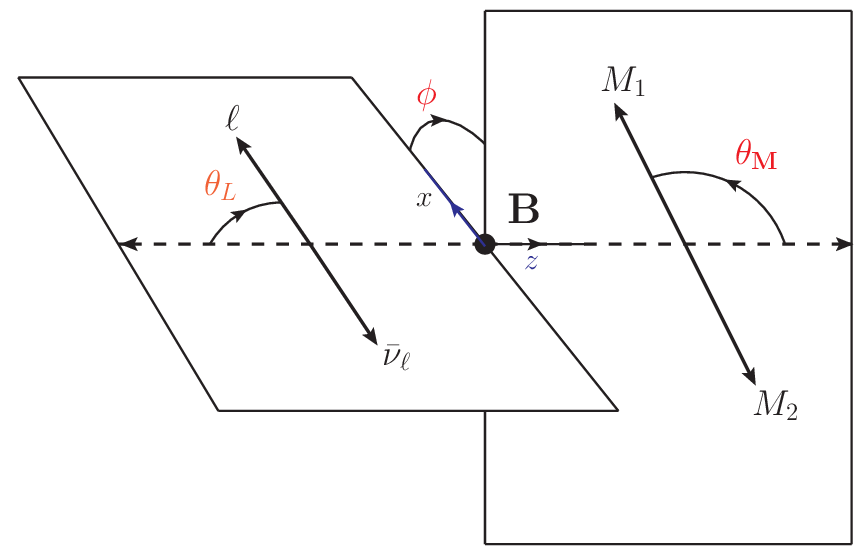}
\caption{Definition of angles $\theta_L$, $\theta_M$ and $\phi$ in the cascade
  decay $D\to M_1M_2 \ell^+\nu_\ell$}
\label{fig:angle}
\end{figure}
First of all, we define kinematic variables and discuss their relations.
A four-body semi-leptonic decay $D\to M_{1}M_{2}\ell^+\nu_\ell$ is
considered, where $D$ is the parent meson, $M_{1(2)}$ is the product meson,
and $\ell=e,\,\mu$. The momentum four-vectors and invariant masses are denoted
by $p$ and $m$, respectively. For convenience, the independent four-vectors
combinations are defined as
\begin{eqnarray}
P^\mu=p^\mu_{M_1}+p^\mu_{M_2},\,\,\,\,Q^\mu=p^\mu_{M_1}-p^\mu_{M_2},\,\,\,\,L^\mu&=&p^\mu_{\ell}+p^\mu_{\nu},\,\,\,\,N^\mu=p^\mu_{\ell}-p^\mu_{\nu},
\end{eqnarray}

A 4-body decay can be uniquely described by kinematically parameterized with
five variables (besides spin). The squared masses of the hadronic system
$M_1M_2$ and the leptonic system $\ell^+\nu_\ell$ are chosen as two of
the five variables,
\begin{eqnarray}
  s_M&=&P^2, \,\,\,\ s_L=L^2,
\end{eqnarray}
and the following relations can be easily derived,
\begin{eqnarray}
  Q^2 =2 m_{M_1}^2+2m_{M_2}^2-s_M,\,\ N^2 =2m_{\ell}^2+2m_{\nu}^2-s_L,
\end{eqnarray}
\begin{eqnarray}
  L\cdot P =\frac{ m_D^2-s_M-s_L}{2},\,\,\,\, L\cdot N &=& m_{\ell}^2-m_{\nu}^2,\,\,\,\, P\cdot Q = m_{M_1}^2-m_{M_2}^2.
\end{eqnarray}

The other three variables are chosen to be: the angle between the $M_2$
three-momentum and the $D$ direction in the $M_1M_2$ rest frame~($\theta_M$);
the angle between the $\nu_\ell$ and the $D$ direction in the
$\ell^+\nu_\ell$ rest frame~($\theta_L$); the angle between the two
decay planes ($\phi$).\footnote{$\phi$ in this paper is defined as $-\phi$
in~\cite{Pais:1968zza,Lee:1992ih}} The angles $\theta_M$, $\theta_L$, $\phi$
are illustrated in Fig.~\ref{fig:angle}.
Various relations of scalar-product invariants can be written as
\begin{eqnarray}
  L\cdot Q &=& L\cdot P\frac{m_{M_1}^2-m_{M_2}^2}{s_M} +X\beta_M \cos\theta_M\,,\nonumber\\
  N\cdot P &=& L\cdot P\frac{m_{\ell}^2-m_{\nu}^2}{s_L} +X\beta_L \cos\theta_L\,,\nonumber\\
   N\cdot Q &=&L\cdot P\beta_M \beta_L \cos\theta_M \cos\theta_L +\frac{m_{\ell}^2-m_{\nu}^2}{s_L}X\beta_M \cos\theta_M+\frac{ m_{M_1}^2-m_{M_2}^2}{s_M}X\beta_L \cos\theta_L\nonumber\\
   &+&L\cdot P\frac{ m_{M_1}^2-m_{M_2}^2}{s_M}\frac{m_{\ell}^2-m_{\nu}^2}{s_L} -\sqrt{s_M}\sqrt{s_L}\beta_M \beta_L \sin\theta_M \sin\theta_L \cos\phi\,,\nonumber\\
   \epsilon_{\mu\nu\rho\sigma}L^{\mu}&N^{\nu}&P^{\rho}Q^{\sigma}= X\sqrt{s_M}\sqrt{s_L}\beta_M \beta_L \sin\theta_M \sin\theta_L \sin\phi\,,
\end{eqnarray}
where $\beta_M$ is the three-momentum modulus of the meson in the
center-of-mass frame of the meson-meson system, $\beta_L$ the three-momentum
modulus of the lepton in the center-of-mass frame of the lepton-neutrino
system, and $X$ an element of phase space,
\begin{eqnarray}
  \beta_M&=&\sqrt{(s_M-m^2_{+})(s_M-m^2_{-})}/s_M \,,\nonumber\\
  \beta_L&=&\sqrt{(s_L-m^2_{L^+})(s_L-m^2_{L^-})}/s_L \,,\nonumber\\
  X&=&\sqrt{m_D^4+s_L^2+s_M^2-2s_Dm_L^2-2s_Mm_D^2-2s_Ms_L}/2 \,,
\end{eqnarray}
with
\begin{eqnarray}
  m_{L^{+}}=m_{\ell}+m_{\nu};\, m_{L^{-}}&=&m_{\ell}-m_{\nu};\,m_{+}=m_{M_1}+m_{M_2};\, m_{-}=m_{M_1}-m_{M_2}.
\end{eqnarray}
Comparing to Refs.~\cite{Pais:1968zza,Lee:1992ih}, we don't neglect the mass of
leptons.

Next, from the effective Hamiltonian at the quark level for
$D\to M_{1}M_{2}\ell^+\nu_{\ell}$, the decay amplitude is given by
\begin{eqnarray}\label{amp1}
  {\cal A}(D\to M_{1}M_{2}\ell^+\nu_{\ell}) &=& \frac{G_F}{\sqrt{2}}V_{q_1q_2}\langle M_{2}M_{1}|\bar{q}_1\gamma_{\mu}(1-\gamma_{5})q_2|D\rangle \bar{u}(p_\ell)\gamma^{\mu}(1-\gamma_{5})v(p_\nu)\,,
\end{eqnarray}
where $G_F$ is the Fermi constant and $V_{q_1q_2}$ is the element of the
Cabibbo-Kobayashi-Maskawa matrix. The hadronic matrix element can be written
in terms of four form factors $w_\pm$, $r$ and $h$ that are defined by
\begin{eqnarray}\label{BtoMM}
 \langle M_{2}M_{1}|\bar{q}_1\gamma_{\mu}(1-\gamma_{5})q_2|D\rangle
&=&
h\epsilon^{\mu\nu\alpha\beta}p_D^\nu P^\alpha Q^\beta+irL^\mu+iw_+ P^\mu+iw_-Q^\mu\,,
\end{eqnarray}
where the form factors $w_\pm$, $r$, and $h$ are function of $s_M$, $s_L$,
and $\cos\theta_M$. The $\epsilon^{\mu\nu\alpha\beta}$ is Levi-Civita symbol.

The differential decay rate takes the form
\begin{eqnarray}
  d\Gamma &=& \frac{G_F^2|V_{q_1q_2}|^2}{(4\pi)^6 m_{D}^3}X\beta_{M}{I}(s_M, s_L, \theta_M, \theta_L, \phi)ds_{M} ds_{L} d{\rm cos} \theta_{M} {\it d}{\rm cos}\theta_{L}d\phi\,.
\end{eqnarray}
In order to study the structure of the hadron system, i.e.~form factors of the
$M_1M_2$ system, the decay intensity $I$ is decomposed with respect to
$\theta_L$ and $\phi$, written as
\begin{eqnarray}
  I &=& I_1+I_2\cos2\theta_L+I_3\sin^2\theta_L\cos2\phi+I_4\sin2\theta_L\cos\phi+I_5\sin\theta_L\cos\phi+I_6\cos\theta_L\nonumber\\
  &&+I_7\sin\theta_L\sin\phi+I_8\sin2\theta_L\sin\phi+I_9\sin^2\theta_L\sin2\phi,
\end{eqnarray}
where $I_{1,...,9}$ depend only on $s_M$, $s_L$, and $\phi$. One can further
express these $I_{1,...,9}$ in terms of form factors:
\begin{eqnarray}
  I_1 &=& \frac{1}{4}(2-\beta_L)\beta_L|F_1|^2 +\left(\frac{\beta_L}{2}-\frac{\beta_L^2}{8}\right)\sin^2\theta_M(|F_2|^2+|F_3|^2)+\frac{1}{2}(1-\beta_L)\beta_L|F_4|^2\,,\nonumber\\
  I_2 &=& -\frac{\beta_L^2}{4}\left[|F_1|^2-\frac{1}{2}\sin^2\theta_M(|F_2|^2+|F_3|^2)\right]\,,\nonumber\\
  I_3 &=& -\frac{\beta_L^2}{4}\left[\sin^2\theta_M(|F_2|^2-|F_3|^2)\right]\,,\nonumber\\
  I_4 &=& \frac{\beta_L^2}{2}\sin\theta_M {\rm Re}(F_1F_2^\star)\,,\nonumber\\
  I_5 &=& -\beta_L\sin\theta_M\left\{{\rm Re}(F_1F_3^\star)-(1-\beta_L){\rm Re}(F_2F_4^\star)\right\}\,,\nonumber\\
  I_6 &=& -\beta_L\sin^2\theta_M {\rm Re}(F_2F_3^\star)-\beta_L(1-\beta_L) {\rm Re}(F_1F_4^\star)\,,\nonumber\\
  I_7&=& \beta_L\sin\theta_M \left\{{\rm Im}(F_1F_2^\star)+(1-\beta_L) {\rm Im}(F_3F_4^\star)\right\}\,,\nonumber\\
  I_8 &=&-\frac{\beta_L^2}{2}\sin\theta_M {\rm Im}(F_1F_3^\star)\,,\nonumber\\
  I_9 &=&\frac{\beta_L^2}{2}\sin^2\theta_M {\rm Im}(F_2F_3^\star)\,,
  \label{eq}
\end{eqnarray}
where $F_{1-4}$ are the form factors,
\begin{eqnarray}
  F_1 &=&X w_++(\beta_M P\cdot L\cos\theta_M+\frac{m_+m_-}{s_M}X)w_-\,,\nonumber\\
  F_2 &=&\beta_M\sqrt{s_M}\sqrt{s_L}w_-\,,\nonumber\\
  F_3 &=&X\beta_M\sqrt{s_M}\sqrt{s_L} h\,,\nonumber\\
  F_4 &=&s_L r+P\cdot Lw_++(X\beta_M\cos\theta_M+\frac{m_+m_-}{s_M}P\cdot L)w_-\,.
  \label{eqF}
\end{eqnarray}

For purpose of discussing angular momentum of $M_1M_2$, e.g.~$S$- and $P$-wave,
the partial wave expansions in spherical harmonics for the form factors
$F_{1-4}$ are written as
\begin{eqnarray}
  F_1(s_M,s_L,\cos\theta_M) &=& \sum_{l=0}^{\infty} F_{1l}(s_M,s_L)P_l(\cos\theta_M)\,,\nonumber\\
  F_2(s_M,s_L,\cos\theta_M) &=& \sum_{l=1}^{\infty}\frac{1}{\sqrt{l(l+1)}} F_{2l}(s_M,s_L)\frac{dP_l(\cos\theta_M)}{d\cos\theta_M}\,,\nonumber\\
  F_3(s_M,s_L,\cos\theta_M) &=& \sum_{l=1}^{\infty}\frac{1}{\sqrt{l(l+1)}} F_{3l}(s_M,s_L)\frac{dP_l(\cos\theta_M)}{d\cos\theta_M}\,,\nonumber\\
  F_4(s_M,s_L,\cos\theta_M) &=&\sum_{l=0}^{\infty} F_{4l}(s_M,s_L)P_l(\cos\theta_M)\,.
\end{eqnarray}

Moreover, the decay of $D\to M_{1}M_{2}\ell^+\nu_\ell$ could happen via
intermediate states, such as vector or scalar mesons, which provides
information about the intermediate resonances~\cite{Wang:2020pyy,Dai:2018vzz}.
The amplitudes of $D\to(S\to M_{1}M_{2})\ell^+\nu_{\ell}$ and
$D\to(V\to M_{1}M_{2})\ell^+\nu_{\ell}$ can be given by
\begin{eqnarray}
{\cal A}(D\to(V\to M_{1}M_{2})\ell^+\nu_{\ell}) &=& \langle V|\bar{q}_1\gamma_{\mu}(1-\gamma_{5})q_2|D\rangle\epsilon\cdot Q g_{VM_{1}M_{2}} D_{FV} \bar{u}(p_\ell)\gamma^{\mu}(1-\gamma_{5})v(p_\nu)\,,\nonumber\\
{\cal A}(D\to(S\to M_{1}M_{2})\ell^+\nu_{\ell}) &=& \langle S|\bar{q}_1\gamma_{\mu}(1-\gamma_{5})q_2|D\rangle g_{SM_{1}M_{2}}D_{FS} \bar{u}(p_\ell)\gamma^{\mu}(1-\gamma_{5})v(p_\nu)\,,
\end{eqnarray}
where
\begin{eqnarray}
  \langle V|\bar{q}_1\gamma_{\mu}(1-\gamma_{5})q_2|D\rangle &=& -\epsilon_{\mu\nu\alpha\beta}\epsilon^{\nu\ast}P_{D}^{\alpha}P^\beta\frac{2V_0(s_L)}{m_V+m_D}-i\left(\epsilon^{\ast}_\mu-\frac{\epsilon^{\ast}\cdot L}{L^2}L_\mu\right)(m_V+m_D)A_1(s_L)\,,\nonumber\\
  &+&i\left({P_{D}}_\mu+P_{\mu}-\frac{m^2_D-m^2_V}{L^2}L_\mu\right)\epsilon^{\ast}\cdot L \frac{A_2(s_L)}{m_V+m_D}-i\frac{2m_V\epsilon^{\ast}\cdot L}{L^2}L_\mu A_0(s_L)\,,\nonumber\\
  \langle S|\bar{q}_1\gamma_{\mu}(1-\gamma_{5})q_2|D\rangle &=& i(f^+(s_L)P_\mu+f^-(s_L)L_\mu)\,,
\end{eqnarray}
with
\begin{eqnarray}
  \sum\epsilon^{\mu\ast}\epsilon^{\nu} &=& -g^{\mu\nu}+\frac{P^\mu P^\nu}{P^2}\,,\nonumber\\
              {\cal A}(D\to M_{1}M_{2}\ell^+\nu_{\ell}) &=& {\cal A}(D\to(V\to M_{1}M_{2})\ell^+\nu_{\ell})+{\cal A}(D\to(S\to M_{1}M_{2})\ell^+\nu_{\ell})\,.
\end{eqnarray}
Here, $f^\pm(s_L)$ is the $D\to S$ form factor; $A_{0,1,2}$ are the 
$D\to V$ axial-vector form factors and $V_0$ is
the $D\to V$ vector form factor~\cite{Ali:1998eb, Cheng:1996if}.
Finally, one can obtain $F_{1-4}$ in the helicity basis (for $S$- and $P$-wave only):
\begin{eqnarray}
  F_1(s_M,s_L,\cos\theta_M) &=& Xf^+(s_L)g_{SM_{1}M_{2}}D_{FS}\nonumber\\
  &+&\cos\theta_M\beta_Mg_{VM_{1}M_{2}} D_{FV}\sqrt{s_Ls_M} H_0(s_L)\,,\nonumber\\
  F_2(s_M,s_L,\cos\theta_M) &=& \frac{1}{2}\beta_Mg_{VM_{1}M_{2}} D_{FV}\sqrt{s_Ls_M}(H_+(s_L)+H_-(s_L))\,,\nonumber\\
  F_3(s_M,s_L,\cos\theta_M) &=& \frac{1}{2}\beta_Mg_{VM_{1}M_{2}} D_{FV}\sqrt{s_Ls_M}(H_+(s_L)-H_-(s_L))\,,\nonumber\\
  F_4(s_M,s_L,\cos\theta_M) &=&s_Lf^-(s_L)g_{SM_{1}M_{2}}D_{FS}\nonumber\\
  &+&2\cos\theta_M\beta_Mg_{VM_{1}M_{2}} D_{FV} \sqrt{s_Ls_M} H_t(s_L)\,,
\end{eqnarray}
with
\begin{eqnarray}
   H_0(s_L) &=& \frac{1}{\sqrt{s_Ls_M}}[P\cdot L(m_V+m_D)A_1(s_L)-2\frac{X^2}{m_V+m_D}A_2(s_L)]\,,\nonumber\\
    H_\pm(s_L) &=& (m_V+m_D)A_1(s_L)\mp\frac{2X}{m_V+m_D}V_0(s_L)\,,\nonumber\\
   H_t(s_L) &=& \frac{X}{\sqrt{s_L}}[A_0(s_L)+A_1(s_L)+A_2(s_L)]\,,
\end{eqnarray}
where $g_{SM_{1}M_{2}}(g_{VM_{1}M_{2}})$ is coupling constant; $D_{FS}(D_{FV})$
is derived from the propagator for $S(V)$. In the case of Breit-Wigner
lineshapes, $D_{FS}=1/(s_M-m^2_S+im_S\Gamma_S)$ and
$D_{FV}=1/(s_M-m^2_V+im_V\Gamma_V)$, or
$D_{FS}=1/(s_M-m^2_S+i\frac{s_M}{m_S}\Gamma_S)$ and
$D_{FV}=1/(s_M-m^2_V+i\frac{s_M}{m_V}\Gamma_V)$ if decay width $\Gamma_{S}(\Gamma_{V})$
is not negligible~\cite{Bohm:2004zi}. The Breit-Wigner lineshape can be
replaced by the Flatt\'e formula~\cite{flatte}, the Gounaris-Sakurai
lineshape~\cite{PhysRevLett.21.244}, etc. for various situations.



\section{Discussion and Conclusion}
In this work, we have derived the parameterization for a four-body
semi-leptonic decay with consideration of the effects caused by the lepton
mass, and expressed the differential decay width in the format used in partial
wave analyses. One can obtain the parameterization formula used in
Refs.~\cite{BESIII:2015hty, BESIII:2018jjm, BESIII:2018qmf} as neglecting the
lepton mass (or setting $\beta_L=1$). While it does not significantly
influence semi-electronic decays, neglecting the lepton mass could lead up to
$\sim1\%$ bias to partial wave analyses for charmed meson semi-muonic decays.

BESIII is accumulating data samples with integrated luminosity of 20~fb$^{-1}$
at center-of-mass energy 3.773~GeV (for $D^0$ and $D^\pm$ mesons) and has
collected 7.33~fb$^{-1}$ data samples at $4.128-4.226$~GeV (for $D_s^+$
mesons)~\cite{BESIII:2020nme}. In addition, the proposed super-tau-charm
factory will be able to reach much enhanced luminosity. These aim for testing
the Standard Model with high precision in the charm sector, but a precise
theoretical parameterization is also needed. With the correction for the
lepton mass presented in this work, partial wave analyses can be performed to
study the form factors in $D_{(s)}\to V\mu^+\nu_\mu$ and
$D_{(s)}\to S\mu^+\nu_\mu$ decays more precisely.

\section*{ACKNOWLEDGMENTS}
HZ and BCK were supported in part by Joint Large-Scale Scientific Facility Fund of the National Natural Science Foundation of China (NSFC) and the Chinese Academy of Sciences under Contracts No.~U2032104 and NSFC under Contracts No.~11875054 and No.~12192263; YY was supported in part by NSFC under Contracts No.~11905023, No.~12047564 and No.~12147102, the Natural Science Foundation of Chongqing (CQCSTC) under Contracts No.~cstc2020jcyj-msxmX0555 and the Science and Technology Research Program of Chongqing Municipal Education Commission under Contracts No.~KJQN202200605 and No.~KJQN202200621.

\end{document}